\documentclass[runningheads]{llncs}
\usepackage[T1]{fontenc}
\usepackage{graphicx}
\usepackage{booktabs}
\usepackage{array}
\usepackage{amsmath}
\usepackage{amssymb}
\usepackage{hyperref}
\usepackage{tabularx}
\usepackage{biblatex}
\usepackage{fontawesome5}
\usepackage{soul}
\begin{document}
\title{Digital Twin Degradation: Detecting Cyber–Physical Attacks via Temporal Inconsistencies}
%
%
\author{Konstantinos E. Kampourakis\inst{1}\orcidID{0009-0000-8883-0735}\faIcon{envelope} \and
Vasileios Gkioulos \inst{1}\orcidID{0000-0001-7304-3835} \and
Sokratis Katsikas \inst{1}\orcidID{0000-0003-2966-9683}}
\authorrunning{Konstantinos E. Kampourakis}
%
\institute{Norwegian University of Science and Technology (NTNU), 2802 Gjøvik, Norway \email{\{konstantinos.kampourakis, vasileios.gkioulos, sokratis.katsikas\}@ntnu.no}}
\maketitle              
\begin{abstract}
Digital Twins (DTs) are increasingly used to monitor and analyze Cyber–Physical Systems (CPS). However, in adversarial environments, the fidelity of a DT cannot be assumed. Communication delays, data manipulation, sensor degradation, or partial information loss may cause the DT state to diverge from the physical process it represents. Such divergence creates temporal inconsistencies that may reveal cyber–physical attacks. This paper proposes a detection framework that monitors temporal consistency between the physical system and a potentially degraded DT view. A DT predictor is trained exclusively on normal system behavior to model short-term system dynamics. During operation, discrepancies between predicted and observed states are transformed into multi-horizon temporal features capturing the magnitude, persistence, and evolution of prediction residuals. An unsupervised density model characterizes normal consistency patterns, while a sequential change detection mechanism identifies sustained deviations indicative of attacks. The approach is evaluated on three widely used Industrial Control System (ICS) datasets, SWaT, HAI, and BATADAL, under multiple DT degradation scenarios, including time desynchronization and partial observability loss. Results show that temporal inconsistency patterns enable reliable event-level attack detection with bounded false alarm rates and low detection latency. The proposed method achieves up to 98\% detection reliability on SWaT and false alarm rates below 2\%. Unlike conventional anomaly detection methods, the proposed framework does not require attack signatures or labeled attack data and remains effective even when the DT view is degraded. These results suggest that DT degradation, often treated as a limitation, can instead serve as a useful signal for cyber–physical security monitoring.

\end{abstract}
\section{Introduction}
\label{S:Intro}

Industrial Cyber-Physical Systems (ICPS) are undergoing a shift toward model-driven monitoring, where predictive representations of system behavior are used alongside raw sensor data. Digital Twins (DTs) are a central component of this trend, providing continuous estimates of expected system dynamics~\cite{Kampourakis2025a}. As their deployment expands across critical infrastructure domains such as water treatment, energy systems, and manufacturing, DTs are increasingly embedded in operational workflows that influence safety and security-critical decisions~\cite{Kampourakis2025}.

Most existing DT-based security approaches implicitly assume that the DT remains a faithful and synchronized representation of the physical system. In practice, however, this assumption is fragile. Industrial environments are subject to communication delays, packet loss, partial observability, sensor faults, and data manipulation~\cite{homaei2024review,Kampourakis2026}. Under adversarial conditions, attackers may intentionally induce such effects to degrade the DT’s view of the system without directly compromising the underlying physical process. As a result, the DT may drift temporally or structurally from the real system, creating discrepancies that traditional anomaly detection mechanisms are not designed to interpret.

This paper argues that DT degradation should not be treated solely as a limitation, but rather as a potential source of security-relevant information. When a DT is trained on normal system dynamics, sustained deviations between its predictions and the observed system state reveal temporal inconsistencies that are difficult to conceal during cyber–physical attacks~\cite{masi2023securing}. These inconsistencies appear not as isolated outliers but as structured temporal patterns whose magnitude, persistence, and evolution encode useful indicators of abnormal system behavior. This reframes degradation from a source of uncertainty into a measurable signal.

Detecting such deviations presents several challenges. First, instantaneous prediction errors are inherently noisy and may arise from benign operational variability. Second, degraded DT views may contain missing or delayed information, complicating direct pointwise comparisons. Third, labeled attack data is rarely available in operational industrial environments, limiting the applicability of supervised detection methods. Effective detection, therefore, requires unsupervised modeling of normal temporal consistency combined with sequential analysis capable of distinguishing transient fluctuations from sustained divergence~\cite{xu2021digital}.

To address these challenges, this work proposes a detection framework that monitors temporal inconsistencies between a physical system and a potentially degraded DT view. A predictive DT model is trained exclusively on normal system behavior to forecast short-term system dynamics. Prediction residuals are transformed into multi-horizon temporal features that capture error magnitude, persistence, and trend across time. An unsupervised density model characterizes normal consistency patterns, while a sequential change detection mechanism identifies sustained deviations with controlled false alarm rates.

The proposed framework is evaluated using three widely studied Industrial Control System (ICS) datasets: SWaT~\cite{goh2016dataset}, HAI~\cite{shin2020hai}, and BATADAL~\cite{taormina2018battle}. Experiments consider multiple DT degradation scenarios, including time desynchronization and partial information loss. Rather than focusing solely on pointwise anomaly scores, the evaluation emphasizes event-level detection capability, detection latency, and false alarm behavior, which are more meaningful for operational security monitoring. Results show that temporal inconsistency patterns enable reliable detection of cyber–physical attacks even when the DT itself is degraded.

In summary, this paper makes the following contributions:

\begin{itemize}
    \item It introduces temporal inconsistency between a physical system and a degraded DT as a detection signal for cyber–physical attack monitoring.
    \item It proposes an unsupervised detection framework combining predictive DTs, multi-horizon temporal inconsistency features, and sequential change detection.
    \item It evaluates the approach on three industrial datasets (SWaT, HAI, BATADAL) under realistic DT degradation scenarios using event-level detection metrics relevant to operational environments.
\end{itemize}

The remainder of this paper is organized as follows. Section~\ref{S:RW} reviews the related work on DTs and anomaly detection in CPS. Section~\ref{S:Prob} formulates the problem, and Section~\ref{S:Method} presents the methodology. Section~\ref{S:Exper} describes the experimental setup and degradation scenarios. Section~\ref{S:Results} reports and analyzes the experimental results. Finally, Section~\ref{S:Disc} discusses limitations and practical implications, and Section~\ref{S:Con} concludes the paper.

\section{Related Work}
\label{S:RW}

In this section, we elaborate on prior research related to anomaly detection in CPS, DT–based security monitoring, and sequential change detection for industrial environments. We explicitly focus on recent experimental publications rather than theoretical ones.

The work in~\cite{yun2022residual} shows that subtle or stealthy anomalies may produce residuals comparable to normal behavior, making threshold-based detection unreliable. To address this, the authors propose N2RE, an anomaly scoring method based on nearest-neighbor distance over residual patterns, capturing similarity rather than magnitude. Experiments on SWaT, WADI, and HAI datasets demonstrate that N2RE improves detection performance by approximately 19\% in F1 score and yields more stable results across models and thresholds. However, the method still depends on residual quality from underlying regression models and requires careful windowing and nearest-neighbor search, which may introduce computational overhead in large-scale settings.

The authors in~\cite{xu2021digital} propose ATTAIN, a DT-driven anomaly detection framework that integrates a timed automaton-based DT model with a GAN-based detector. The DT model is continuously updated using both historical and real-time data, enabling it to simulate CPS behavior and generate ground truth labels. These labels are then used to supervise a semi-supervised GAN that distinguishes between normal and anomalous system states. The approach is evaluated on SWaT, WADI, and BATADAL datasets, showing consistent improvements over baseline methods such as LSTM-CUSUM and MAD-GAN in terms of F1 score and detection performance. However, the method introduces architectural complexity, relies on accurate DT modeling, and incurs computational overhead due to GAN training and online DT updates.

Kim et al.~\cite{kim2021revitalizing} propose SOMAD, a CPS anomaly detection framework that leverages forecasting error patterns rather than raw sensor values. The approach constructs structured error representations from sequence models such as seq2seq, MDN, and RNN, and learns normal behavior using a Self-Organizing Map (SOM). Anomalies are identified through statistical hypothesis testing on distances between observed error patterns and learned normal prototypes. Evaluated on SWaT and HAI datasets, the method demonstrates improved time-series-aware detection performance compared to static thresholding and CUSUM baselines. However, the approach remains dependent on the quality of the underlying forecasting models and introduces additional complexity through SOM training and statistical testing mechanisms.

The contribution in~\cite{castellani2020real} investigates the integration of DT-generated data with weakly supervised learning for anomaly detection in CPS. The approach leverages a DT to generate large-scale normal-operation data, which is combined with a small set of labeled anomalies from real systems. Two main methods are proposed: a clustering-based approach with a penalty term and a Siamese Autoencoder (SAE) that enforces separation between normal and anomalous patterns. Experimental results on real-world facility monitoring data show that weakly supervised DT-based methods outperform both unsupervised and fully data-driven baselines, while significantly reducing false positives. The work highlights the practical value of DTs for addressing data scarcity and improving anomaly detection robustness in real industrial environments.

Li et al.~\cite{li2019mad} propose MAD-GAN, an unsupervised deep learning framework that models multivariate CPS data. The method captures temporal dependencies and inter-variable correlations by jointly training a generator and discriminator, and introduces a combined anomaly score (DR-score) that integrates reconstruction error and discrimination output. Evaluated on SWaT and WADI datasets, MAD-GAN demonstrates strong recall and competitive performance compared to traditional and deep learning baselines, particularly in detecting complex cyber-attacks. However, the approach exhibits instability across training iterations and suffers from low precision, highlighting limitations in robustness and threshold-based decision mechanisms.

The authors in~\cite{kravchik2018detecting} propose an anomaly detection approach for ICS based on multivariate time-series prediction using deep neural networks. Their method models normal system behavior on the SWaT dataset and detects anomalies via statistical deviations between predicted and observed sensor values, effectively using normalized residual errors and temporal thresholding. The study demonstrates that 1D CNNs outperform recurrent architectures such as LSTMs in both detection performance and computational efficiency, achieving detection of 31 out of 36 attacks. However, the approach remains fundamentally based on pointwise prediction error and thresholding, without explicitly modeling temporal inconsistency or system-level desynchronization effects across time horizons.

In contrast to prior work, existing approaches primarily rely on either data-driven anomaly detection or DT-assisted data augmentation, without explicitly modeling the temporal inconsistency between the physical system and its DT as the core detection signal. Methods such as SOM-based detection or weakly supervised SAE architectures focus on learning representations of normality and identifying deviations, often requiring either accurate forecasting models, latent space separation, or labeled anomalies. Similarly, DT-based approaches typically use simulated data to improve training coverage or reduce data scarcity, but they do not exploit discrepancies between synchronized DT and real-time observations as an explicit detection mechanism.

The proposed framework, however, differs in that it formulates anomaly detection as a sequential change detection problem over DT inconsistency signals, rather than a static classification or reconstruction task. Instead of learning normal behavior solely from data, the method continuously compares real system observations with a DT-generated reference and aggregates deviations over time using a sequential detector. This enables robust event-level detection, reduced sensitivity to pointwise noise, and resilience to partial DT degradation. As a result, the approach shifts the focus from feature learning and reconstruction accuracy to temporal persistence of inconsistencies, providing a more operationally meaningful and deployment-oriented solution for CPS security monitoring.

\section{Problem Formulation \& Threat Model}
\label{S:Prob}

This section defines the detection problem and the associated threat model. We examine CPS in which a predictive DT operates under imperfect, delayed, or incomplete observations.

First, we consider a CPS consisting of physical processes, sensors, and actuators that generate multivariate time-series data. Let
\[
\mathbf{x}_t \in \mathbb{R}^{n}
\]
denote the vector of sensor measurements at discrete time $t$, and
\[
\mathbf{u}_t \in \mathbb{R}^{m}
\]
denote the corresponding actuator states or control inputs.

A DT is implemented as a predictive model $f_\theta(\cdot)$ trained exclusively on data collected during normal system operation. Given a window of past observations and control inputs of length $K$, the DT predicts future system states as

\[
\hat{\mathbf{x}}_{t+h} = f_\theta\left(\mathbf{x}_{t-K:t-1}, \mathbf{u}_{t-K:t-1}\right),
\]

where $h$ denotes the prediction horizon.

During deployment, predicted values are compared with observed measurements to compute prediction residuals:
\[
\mathbf{r}_{t+h} = \mathbf{x}_{t+h} - \hat{\mathbf{x}}_{t+h},
\] 

These residuals quantify discrepancies between predicted system evolution and actual system behavior.

As already mentioned in Section~\ref{S:Intro}, conventional DT-based monitoring approaches implicitly assume that the DT maintains a synchronized and complete view of the physical system. In practice, this assumption may not hold. Industrial environments often experience communication delays, packet loss, and incomplete sensor observations. Therefore, we explicitly consider scenarios in which the DT receives degraded information. DT degradation is modeled as distortions in the information available to the predictive model rather than modifications to the physical process itself. Specifically, degraded DT views arise from delayed sensor updates, intermittent data loss, and partial observability caused by communication disruptions. Under such conditions, the DT continues to generate predictions based on outdated or incomplete observations, introducing temporal misalignment between the DT state and the true system state. 

It is important to distinguish DT degradation from cyber-physical attacks affecting the underlying process. In this work, DT degradation refers to imperfections in the information available to the DT, such as delayed updates, packet loss, or incomplete observations, while the physical process itself continues to evolve normally. Under benign degradation, inconsistencies arise primarily from temporal misalignment and incomplete visibility.

In contrast, cyber-physical attacks directly influence the physical process, sensor measurements, actuator behavior, or control logic, producing deviations that alter the true system dynamics. These attacks may occur independently or simultaneously with DT degradation. In the combined case, the DT receives delayed or incomplete information while the physical system is also being manipulated adversarially. This creates compound inconsistencies that reflect both synchronization errors and attack-induced deviations.

The proposed framework does not attempt to explicitly classify whether inconsistencies originate from benign degradation, adversarial manipulation, or both simultaneously. Instead, the detector focuses on identifying persistent temporal inconsistencies between DT predictions and observed system behavior. Sequential aggregation is then used to distinguish sustained abnormal deviations from transient inconsistencies caused solely by moderate benign degradation.

In the absence of attacks, prediction errors typically remain bounded and exhibit relatively stable temporal behavior under moderate DT degradation. In contrast, cyber-physical attacks induce persistent deviations that cannot be explained by benign degradation alone.

Rather than relying solely on instantaneous residual magnitudes, we analyze temporal inconsistencies that evolve across time and across multiple prediction horizons. For each horizon $h$, residuals are summarized using norm-based statistics, and multi-horizon features are constructed to capture the persistence of prediction errors, growth trends across prediction horizons, and temporal variability of residual magnitudes. 

Let $\mathbf{z}_t \in \mathbb{R}^d$ denote the resulting feature vector at time $t$, derived from residual statistics across multiple horizons and auxiliary indicators such as delay magnitude and invariant violations. Under normal system operation, $\mathbf{z}_t$ follows a stationary distribution learned from normal data. The detection problem can therefore be formulated as identifying time intervals in which the distribution of $\mathbf{z}_t$ deviates persistently from its nominal behavior.

Our key objective is to detect attack events as early as possible while maintaining a low false alarm rate during normal operation. Formally, given a sequence of feature vectors $\{\mathbf{z}_t\}$, the detector produces a binary alarm signal $a_t \in \{0,1\}$ such that

\begin{itemize}
    \item false alarms during normal operation remain bounded,
    \item detection delay for sustained attacks is minimized,
    \item transient fluctuations and noise do not trigger alarms.
\end{itemize}

To achieve this, anomaly scores derived from an unsupervised density model are processed using a sequential change detection mechanism. This temporal aggregation enables the detector to distinguish sustained deviations from isolated outliers and to operate reliably under degraded DT conditions.

Regarding threat modeling, we assume an adversary capable of launching cyber-physical attacks that alter system behavior through manipulation of sensors, actuators, or control logic. The attacker may also disrupt communication channels between the physical system and the DT, resulting in delayed or missing observations. Specifically, the attacker may attempt to induce gradual or stealthy deviations in system dynamics, evade simple threshold-based anomaly detectors, or exploit communication disruptions to degrade the DT view of the system.

We also assume that labeled attack data is unavailable during training and that the DT is trained solely on normal operation data. The attacker is not assumed to have direct access to or control over the DT model parameters. Detection is therefore based on monitoring temporal inconsistencies between the physical system and the potentially degraded DT predictions.

\section{Methodology}
\label{S:Method}

In this section, we present the proposed detection framework, which combines a predictive DT, multi-horizon residual analysis, unsupervised density modeling, and sequential change detection to identify cyber–physical attacks under degraded DT conditions.

Specifically, the DT is implemented as a sequence-to-one predictor trained exclusively on normal operation data. Given a sliding window of length $K$, the model learns the mapping
\[
f_\theta : \left(\mathbf{x}_{t-K:t-1}, \mathbf{u}_{t-K:t-1}\right) \rightarrow \hat{\mathbf{x}}_{t},
\]
by minimizing the mean squared prediction error
\[
\mathcal{L}(\theta) = \mathbb{E}\left[\lVert \mathbf{x}_{t} - \hat{\mathbf{x}}_{t} \rVert_2^2 \right].
\]
Since the model is trained only on normal data, it captures nominal system dynamics and serves as a reference during deployment.

To capture temporal effects beyond one-step prediction, the DT is rolled forward to produce predictions at multiple horizons $h \in \mathcal{H}$:
\[
\hat{\mathbf{x}}_{t+h}, \quad h \in \{1,3,5,10\}.
\]
Prediction residuals are computed as
\[
\mathbf{r}_{t+h} = \mathbf{x}_{t+h} - \hat{\mathbf{x}}_{t+h},
\]
representing discrepancies between predicted and observed system states. These residuals capture the accumulation of mismatch between the DT and the physical process over time.

Residuals across all horizons are summarized into a feature vector $\mathbf{z}_t \in \mathbb{R}^d$. The feature representation includes norm-based statistics such as $\ell_2$ and $\ell_\infty$ residual magnitudes for each horizon, inter-horizon trend features capturing error growth, and rolling statistics of short-horizon residuals. Additional indicators, such as delay magnitude and invariant violation counts, are incorporated to reflect degradation effects. Missing or invalid features caused by degraded inputs are handled using median imputation estimated from normal training data, ensuring a fixed-dimensional representation.

An unsupervised density model is trained on feature vectors derived from normal operation. Let $\mathcal{Z}_{\text{train}}$ denote the set of valid normal feature vectors. The model estimates the likelihood of observing a feature vector $\mathbf{z}_t$, and defines an anomaly score
\[
s_t = -\log p(\mathbf{z}_t),
\]
where larger values indicate stronger deviation from nominal behavior. This step reduces the high-dimensional feature representation to a univariate anomaly score while preserving sensitivity to distributional changes.

Because isolated anomaly scores may arise from noise or benign DT degradation, temporal aggregation is applied. Scores are standardized using statistics estimated from normal validation data:
\[
z_t = \frac{s_t - \mu}{\sigma}.
\]
A two-sided Cumulative Sum (CUSUM) procedure is then used:
\[
g_t^{+} = \max\left(0, g_{t-1}^{+} + (z_t - k)\right), \quad
g_t^{-} = \min\left(0, g_{t-1}^{-} + (z_t + k)\right).
\]
An alarm is triggered when
\[
\max(g_t^{+}, -g_t^{-}) > h,
\]
where $k$ is the reference value and $h$ is the detection threshold. To reduce fragmentation, alarms are extended using a hold interval followed by a cooldown period that suppresses immediate retriggering.

\begin{figure}
    \includegraphics[width=\linewidth]{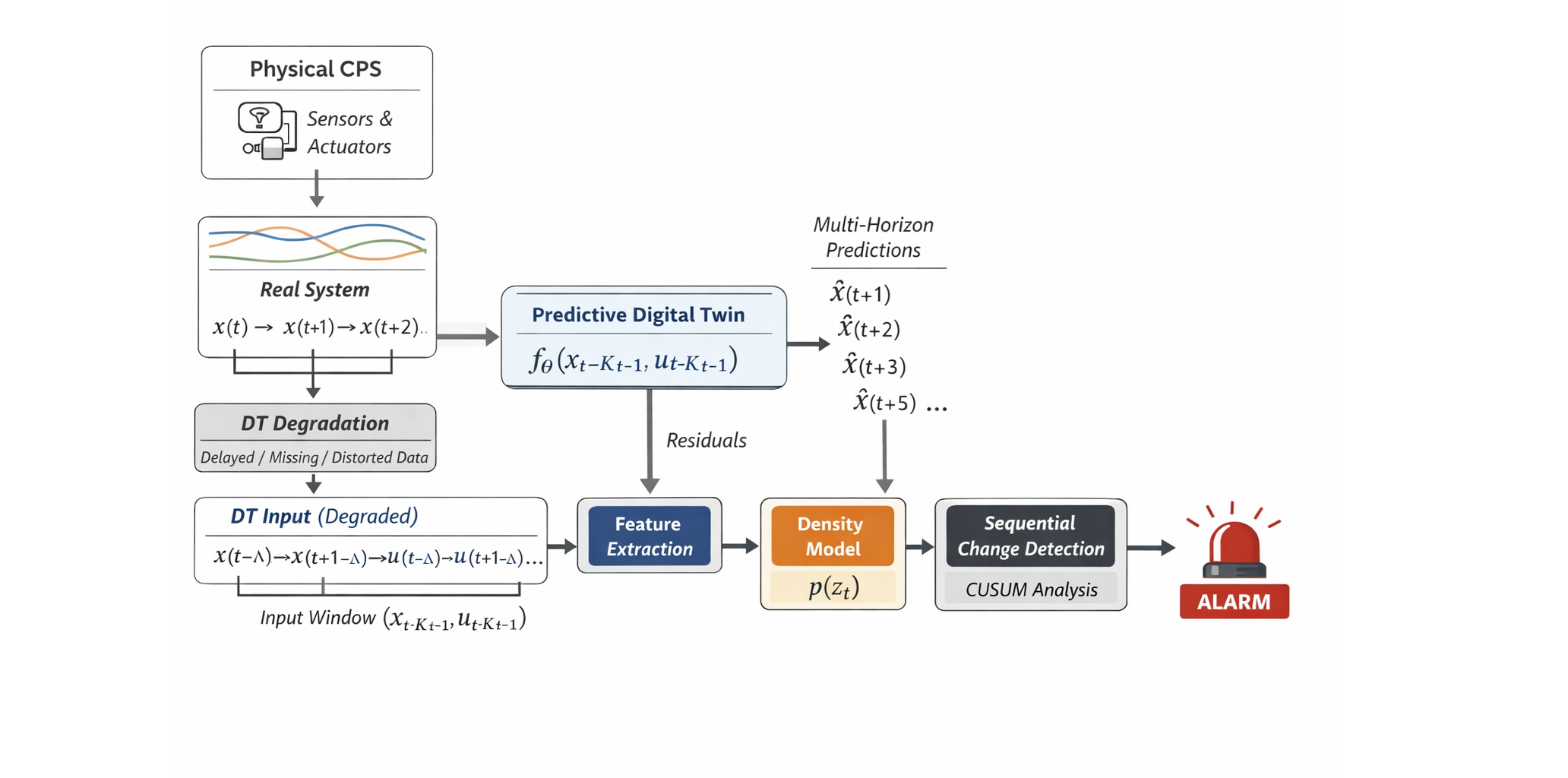}
    \caption{Digital Twin inconsistency detection pipeline.}
    \label{F:Framework}
\end{figure}

Overall, the framework separates modeling, feature extraction, scoring, and decision-making into distinct stages. As shown in Figure~\ref{F:Framework}, the physical CPS generates sensor and actuator data that define the true system evolution. A predictive DT, trained exclusively on normal operation data, receives a potentially degraded version of this information due to delay, missing observations, or distortion. Using the degraded input window, the DT produces multi-horizon predictions of future system states. These predictions are compared with the corresponding ground truth measurements to compute residuals across multiple prediction horizons. The residuals are then aggregated into a feature vector that captures their magnitude, temporal persistence, and evolution. An unsupervised density model, trained on normal data, assigns anomaly scores to the feature vectors. These scores are subsequently processed by a sequential change detection mechanism, which accumulates evidence over time to distinguish sustained deviations from transient noise. When the accumulated deviation exceeds a threshold, an alarm is triggered, indicating a potential cyber–physical attack.

\section{Experimental Setup}
\label{S:Exper}

This section describes the datasets, DT configuration, degradation models, and evaluation protocol used to validate the proposed detection framework. As already discussed, experiments are conducted on three widely used ICS datasets: SWaT, HAI, and BATADAL.

Particularly, SWaT is a water treatment testbed dataset containing synchronized measurements from sensors and actuators under normal operation and multiple attack scenarios. The HAI dataset contains operational data from an industrial control environment with labeled cyber–physical attack events affecting sensors and control logic. Finally, BATADAL is a benchmark dataset derived from a water distribution system simulation and includes several attack scenarios targeting system integrity.

For all datasets, the system state at time $t$ is represented by sensor measurements
\[
\mathbf{x}_t \in \mathbb{R}^{n_s}
\]
and actuator states
\[
\mathbf{u}_t \in \mathbb{R}^{n_a}.
\]

Binary labels indicate whether the system is under attack. Importantly, only normal data are used for model training. The normal portion of each dataset is divided chronologically into disjoint subsets, with 80\% used for DT training and the remaining 20\% reserved for validation and detector calibration. The split preserves temporal ordering and avoids information leakage between subsets. Sliding windows are constructed independently within each subset, ensuring that no window spans the boundary between training and validation segments. Attack samples are excluded entirely from training and validation and are used only for final evaluation. This setup reflects realistic deployment conditions where attack signatures are unavailable during training.

The DT is implemented as a predictive LSTM model with two recurrent layers and a hidden dimension of 128. The input consists of a sliding window of length $K = 50$ containing both sensor and actuator histories. The model predicts the next sensor state:

\[
\hat{\mathbf{x}}_{t} = f_\theta(\mathbf{x}_{t-K:t-1}, \mathbf{u}_{t-K:t-1}).
\]

Training minimizes mean squared prediction error using the Adam optimizer. Early stopping based on validation loss is applied to prevent overfitting.

To emulate realistic deployment conditions, controlled degradation is applied to the DT input stream during inference. The DT receives delayed observations, simulating communication latency or clock drift. In the high-severity configuration used in the experiments, a fixed delay of 10 time steps is introduced. In addition, observations may be intermittently missing, representing packet loss or partial observability. Importantly, degradation affects only the information available to the DT. Ground truth physical measurements remain unchanged.

Multi-horizon residual features are computed as described in Section~\ref{S:Method}. Residuals are generated for prediction horizons $h \in \{1,3,5,10\}$ and summarized into temporal feature vectors capturing residual magnitude, growth across horizons, and short-term residual statistics. Because degraded DT views may produce missing predictions, feature vectors may contain invalid values. A robust median imputer fitted on normal training features is applied to ensure a fixed-dimensional feature representation.

An unsupervised Gaussian Mixture Model (GMM) with four components and diagonal covariance is trained on imputed feature vectors derived from normal operation. A GMM is selected because it provides a flexible probabilistic representation of normal feature distributions without requiring labeled attack data. Diagonal covariance is used to improve numerical stability and reduce overfitting, given the relatively small dimensionality of the feature space. Anomaly scores are defined as negative log-likelihoods:

\[
s_t = -\log p(\mathbf{z}_t).
\]

Scores are standardized using statistics estimated from normal validation data. A two-sided CUSUM detector is then applied to identify sustained deviations. The reference value $k$ is fixed while the threshold $h$ is calibrated on validation data to achieve a target false alarm rate. Additional parameters include a hold interval of 10 time steps and a cooldown period of 50 steps.

Moreover, performance is evaluated using event-level metrics:

\begin{itemize}
\item \textbf{Segment detection rate (SegDet)}: The number of attack segments that are successfully detected over the total number of segments. A segment is considered detected if at least one alarm is raised within its duration.

\item \textbf{Event recall}: The fraction of attack events that are detected, computed as the ratio of detected segments to total segments.

\item \textbf{False alarm onset rate (FPR)}: The frequency at which the detector incorrectly enters an alarm state during normal operation.

\item \textbf{Mean detection delay}: The average time between the start of an attack segment and the first alarm.

\item \textbf{Reliability}: A combined metric defined as $\text{Recall} \times (1 - \text{FPR})$, capturing the trade-off between detection coverage and false alarm rate.
\end{itemize}

Event-level metrics are particularly important for industrial monitoring because attacks typically span extended time intervals, and operational relevance depends on detecting the event rather than individual anomalous samples.

For each dataset, the DT is trained once using normal operation data. Feature extraction, density modeling, and CUSUM calibration are performed using clean DT inputs only. The trained detector is then evaluated on both clean and degraded DT views without retraining. This protocol isolates the effect of DT degradation and measures the ability of the framework to detect attacks under realistic deployment conditions.

To ensure reproducibility, all components of the proposed framework are defined using fixed configurations and trained exclusively on normal operation data. The DT is implemented as a two-layer LSTM with hidden dimension 128 and input window length $K=50$. The model is trained using the Adam optimizer with early stopping based on validation loss. No attack data is used during training or model selection. Multi-horizon predictions are generated for horizons $\mathcal{H} = \{1,3,5,10\}$. Residuals are computed at each horizon and transformed into feature vectors consisting of norm-based statistics ($\ell_2$, $\ell_\infty$), inter-horizon trends, and rolling statistics. Additional indicators capturing delay magnitude and invariant violations are included. Missing values arising from degraded observations are handled using median imputation fitted on normal training data. The anomaly scoring model is a GMM with four components and diagonal covariance, trained on feature vectors derived from normal operation only. Anomaly scores are computed as negative log-likelihoods. Sequential detection is performed using a two-sided CUSUM procedure. The normalization parameters $(\mu, \sigma)$ are estimated from validation scores, while the remaining parameters are fixed across all experiments as $k=0.05$, $h=3.0$, hold$=10$, and cooldown$=50$. No tuning is performed using attack data. All datasets are processed using consistent feature extraction, model training, and evaluation pipelines. The same hyperparameters are applied across SWaT, HAI, and BATADAL to ensure comparability and to evaluate generalization under different system dynamics.

\section{Results}
\label{S:Results}

This section presents the experimental results of the proposed DT inconsistency detection framework. Performance is evaluated under both clean and degraded DT views using the protocol described in Section~\ref{S:Exper}. Results are reported primarily using event-level metrics that capture detection persistence and delay.

We first evaluate the method on the SWaT dataset. Under a clean DT view, the detector achieves perfect event detection, identifying all attack segments in the dataset (SegDet = 35/35, Recall = 1.00). Detection delay remains low ($\sim$22 time-steps), and the FPR remains below 2\% during normal operation. When temporal desynchronization is introduced, event detection performance remains unchanged (35/35, Recall = 1.00). Detection delay increases slightly to $\sim$26 time-steps, and the FPR increases marginally to 1.65\%, while maintaining high reliability (0.984). These results show that the proposed framework maintains detection performance under DT degradation, with only minor increases in delay and FPR.

\begin{figure}
    \includegraphics[width=\linewidth]{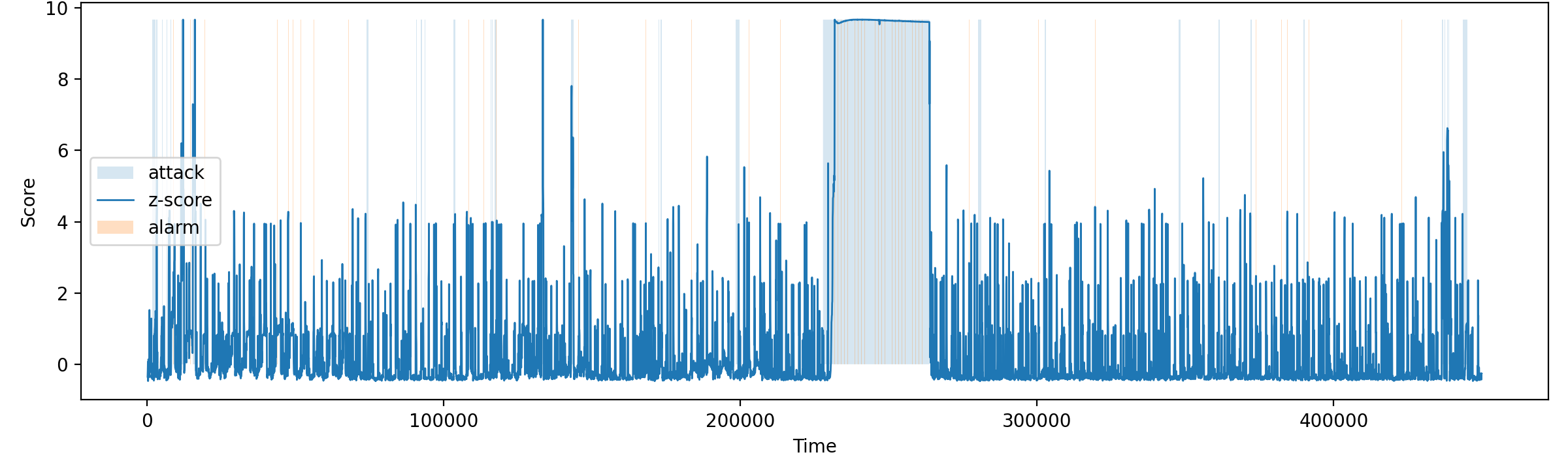}
    \caption{Full detection timeline on the SWaT dataset under clean DT conditions.}
    \label{F:Timeline}
\end{figure}

Figure~\ref{F:Timeline} illustrates the full detection timeline on the SWaT dataset under clean DT conditions. The standardized anomaly score exhibits frequent short-term fluctuations during normal operation, reflecting measurement noise and minor modeling inaccuracies. These fluctuations, however, do not lead to persistent alarm activation due to the sequential nature of the detector. In contrast, during attack intervals, the anomaly score increases and remains elevated over extended periods. This sustained deviation results in the accumulation of evidence in the CUSUM statistic, eventually triggering an alarm. The delay between the onset of the attack and the activation of the alarm corresponds to the time required for this accumulation process, which is consistent with the persistence-based design of the detector. The figure also shows that alarms are not triggered by isolated spikes in the anomaly score, but rather by continuous deviations over time. At the same time, it justifies the strong event-level detection performance, since attack segments are consistently identified once sufficient temporal evidence is present. Overall, the full-series view highlights the trade-off between sensitivity and reliability: the detector avoids spurious alarms during normal operation while maintaining the ability to detect sustained cyber-physical attacks with bounded delay.

We next evaluate the method on the HAI dataset, which contains a larger number of attack events and more complex system dynamics. Under clean DT conditions, the detector identifies nearly all attack segments (SegDet = 48/50, Recall = 0.96) with low FPR (0.48\%) and reliability (0.955). Detection delay remains moderate ($\sim$28 time-steps). When DT desynchronization is introduced, detection performance slightly improves (SegDet = 49/50, Recall = 0.98), while delay increases to $\sim$44 time-steps and FPR rises to 1.67\%. Despite this trade-off, reliability remains high (0.964), indicating robust performance under degraded conditions.

Finally, we evaluate the framework on the BATADAL dataset, which contains fewer attack events and different system dynamics. Under clean DT conditions, the detector identifies 2 out of 5 attack segments (Recall = 0.40) with low FPR (0.50\%) and lower reliability (0.396). When DT desynchronization is introduced, detection performance improves significantly (SegDet = 4/5, Recall = 0.80), while delay increases slightly to $\sim$12 time-steps and FPR rises to 1.67\%. This improvement suggests that DT degradation can amplify attack-induced inconsistencies, making them more detectable in certain scenarios.

The summary of the results is presented in Table~\ref{T:Results}. These results can be explained by the temporal nature of the detection mechanism. The DT captures nominal system dynamics, and prediction errors remain bounded under normal conditions, even when the DT view is partially degraded. In contrast, cyber–physical attacks introduce structured deviations that persist across time and across prediction horizons. While individual residuals may be noisy, their temporal aggregation leads to consistent divergence that is effectively captured by the sequential detector, resulting in strong event-level detection performance with controlled FPR.

\begin{table}[t]
\centering
\begin{tabular}{lcccccc}
\hline
Dataset & Condition & SegDet & Recall & Delay (Time-Steps) & FPR & Reliability \\
\hline
SWaT & Clean & 35/35 & 1.00 & $\sim$22 & 1.38\% & 0.986 \\
     & Degraded & 35/35 & 1.00 & $\sim$26 & 1.65\% & 0.984 \\
\hline
HAI  & Clean & 48/50 & 0.96 & $\sim$28 & 0.48\% & 0.955 \\
     & Degraded & 49/50 & 0.98 & $\sim$44 & 1.67\% & 0.964 \\
\hline
BATADAL & Clean & 2/5 & 0.40 & $\sim$10 & 0.50\% & 0.396 \\
        & Degraded & 4/5 & 0.80 & $\sim$12 & 1.67\% & 0.796 \\
\hline
\end{tabular}
\caption{Event-level detection performance under clean and degraded DT conditions.}
\label{T:Results}
\end{table}

Across all datasets, event-level metrics provide a more meaningful assessment of detection capability than pointwise classification metrics. Cyber–physical attacks typically persist over extended time intervals, and operational response depends on detecting the presence of an attack event rather than identifying every anomalous sample. The proposed framework consistently detects the majority of attack segments across datasets while maintaining bounded FPR during normal operation. Detection delays remain small relative to attack durations, indicating that temporal inconsistencies accumulate quickly once attacks begin to influence system dynamics. This is particularly important in industrial settings, where operators respond to sustained abnormal behavior rather than individual anomalous samples.

To further position the proposed method relative to existing approaches, Table~\ref{T:Comparison} provides a qualitative comparison with representative ICS anomaly detection techniques. The comparison highlights three key differences. First, most existing approaches rely on pointwise residual evaluation or latent representation learning and do not explicitly model the relationship between the physical system and a DT. As a result, they treat prediction errors as independent anomaly signals rather than structured temporal inconsistencies. Second, while some methods incorporate temporal information implicitly, for example, through sequence models or smoothing, they do not employ explicit sequential change detection mechanisms. In contrast, the proposed framework uses a CUSUM-based detector to accumulate evidence over time, enabling robust discrimination between transient fluctuations and sustained deviations. Finally, existing approaches are primarily evaluated using pointwise metrics, which do not directly reflect operational requirements in industrial environments. The proposed method is explicitly designed for event-level detection, prioritizing reliable identification of attack segments with controlled FPR and bounded detection delay. This combination of DT-aware modeling, sequential aggregation, and event-level focus differentiates the proposed framework from prior work.

\begin{table}[t]
\centering
\begin{tabular}{lccc}
\toprule
Method & DT-aware & Sequential Detection & Event-level Focus \\
\midrule
CNN Residual Thresholding & $\times$ & $\times$ & $\times$ \\
MAD-GAN & $\times$ & $\times$ & $\triangle$ \\
SOMAD & $\times$ & $\triangle$ & $\triangle$ \\
Predictor + CUSUM & $\times$ & $\checkmark$ & $\triangle$ \\
Proposed Framework & $\checkmark$ & $\checkmark$ & $\checkmark$ \\
\bottomrule
\end{tabular}
\caption{Qualitative comparison with representative ICS anomaly detection methods. $\checkmark$ indicates full support, $\times$ indicates absence, and $\triangle$ denotes partial capability.}
\label{T:Comparison}
\end{table}

Overall, the experimental results demonstrate that:

\begin{itemize}
\item Temporal inconsistencies between the physical system and a DT provide a reliable signal for detecting cyber–physical attacks.
\item The proposed detection framework maintains strong event-level detection performance on SWaT and HAI, with dataset-dependent behavior on BATADAL.
\item DT degradation does not prevent attack detection and may, in some case,s increase attack observability.
\item Sequential aggregation of anomaly scores enables low FPR while preserving early detection capability.
\end{itemize}

These results suggest that monitoring DT inconsistencies can provide a practical and robust mechanism for cyber–physical security monitoring in industrial environments.

\section{Discussion}
\label{S:Disc}

In this section, we discuss the implications of the experimental results, clarify the intended scope of the proposed approach, and outline current limitations.

The proposed framework is designed for event-level detection rather than precise timestamp labeling. In industrial environments, the primary objective is to identify the presence of abnormal system behavior quickly and reliably, rather than to classify individual samples. From this perspective, event detection rates, false alarm behavior, and detection delays provide a more meaningful assessment of performance~\cite{yu2024leveraging}. Across all evaluated datasets, the framework detects the majority of attack segments while maintaining low FPR during normal operation. Detection delays remain small relative to attack durations, indicating that inconsistencies are rapidly accumulated once they emerge.

This behavior follows directly from the design of the detection pipeline. The DT is trained exclusively on normal operation data, and anomaly detection is performed using an unsupervised density model combined with a sequential change detection mechanism. The use of CUSUM enforces temporal persistence, requiring sustained deviations before raising an alarm. As a result, the detector suppresses transient fluctuations and noise, focusing instead on consistent deviations that reflect structural changes in system behavior. We prioritize temporal consistency to improve operational reliability and reduce spurious alarms, which is critical in industrial control environments where false positives are costly~\cite{choi2024detecting}.

An important observation from the experiments is that moderate DT degradation does not prevent attack detection. In several cases, temporal desynchronization even increases the visibility of attack-induced inconsistencies. This effect arises because attacks typically introduce structured deviations in system dynamics that persist across time. In contrast, benign degradation effects such as delay or missing observations tend to produce short-lived fluctuations. Sequential aggregation allows the detector to distinguish between these two behaviors. However, degradation does not universally improve detection performance. In some datasets, it increases detection delay or time spent in alarm. These results suggest that DT degradation alters the observability of inconsistencies rather than uniformly strengthening detection.

The proposed approach assumes an attacker capable of manipulating sensors, actuators, or communication timing. Such attacks can alter system dynamics or degrade the DT view through delayed or missing observations. We do not assume that the attacker can directly modify the DT model itself. If both the physical process and the DT are simultaneously compromised, detection becomes significantly more difficult. This scenario corresponds to a full DT compromise and lies outside the scope of this work. The method also assumes that attacks persist for multiple time steps, allowing inconsistencies to accumulate. Extremely short perturbations affecting only a single sample are unlikely to trigger alarms.

However, several limitations should be acknowledged. The DT is predictive rather than causal, which limits the interpretability of individual alarms. Detection performance depends on the quality and representativeness of normal training data. The degradation scenarios considered primarily involve temporal desynchronization and partial observability. Other forms of DT degradation, like structural model drift, remain to be studied. While the framework was evaluated on multiple industrial datasets, additional validation on heterogeneous cyber-physical systems would further strengthen the generality of the approach.

The proposed method is not intended to replace traditional intrusion detection systems. Instead, it provides a complementary monitoring layer based on inconsistencies between the physical system and its DT. Unlike approaches based on protocol analysis, attack signatures, or control logic semantics, the framework operates directly on system dynamics. Its main contribution lies in exploiting the relationship between the physical process and its predictive DT to identify abnormal behavior.

\section{Conclusion}
\label{S:Con}

This paper presented a DT–driven framework for detecting cyber–physical attacks by monitoring temporal inconsistencies between a physical system and a potentially degraded DT view. Contrary to traditional intrusion detection approaches that mostly rely on attack signatures, protocol semantics, or supervised learning, the proposed method exploits prediction residuals and their temporal structure as the primary detection signal.

A predictive DT is trained exclusively on normal system behavior. During deployment, multi-horizon prediction residuals are transformed into temporal feature representations capturing the magnitude, persistence, and evolution of inconsistencies. These features are modeled using an unsupervised GMM, while a sequential CUSUM detector identifies sustained deviations with controlled false alarm rates.

Experimental evaluation on three widely used ICS datasets, SWaT, HAI, and BATADAL, demonstrates that the proposed framework consistently detects the majority of attack events while maintaining low FPR during normal operation. Moreover, event-level metrics show that detection delays remain small relative to attack durations, confirming the suitability of the approach for operational monitoring environments.

An important observation is that moderate DT degradation does not prevent attack detection. In several scenarios, temporal desynchronization increases the visibility of attack-induced inconsistencies, highlighting that imperfect DT synchronization can still provide useful security signals.

Overall, the results indicate that monitoring inconsistencies between a physical system and its DT can provide a practical and complementary mechanism for cyber-physical security monitoring. Beyond their traditional roles in simulation and optimization, temporal inconsistencies between DT predictions and real system observations can serve as a useful signal for cyber-physical attack detection, even under partially degraded observation conditions.

Future work will investigate adaptive degradation modeling and improved interpretability of detected inconsistencies. In addition, the framework will be evaluated across a broader range of CPS to assess generalization under different system dynamics and operating conditions.

\subsection*{Acknowledgments}
This work is supported by the Research Council of Norway through
the SFI Norwegian Centre for Cybersecurity in Critical Sectors (NORCICS) project no. 310105
%
%
%
%
\printbibliography
\end{document}